# GRAZSCHUMPETERCENTRE

## GSC Discussion Paper Series

Paper No. 26

# Endogenous viral mutations, evolutionary selection, and containment policy design

Patrick Mellacher[1]

[1] University of Graz, Graz Schumpeter Centre,

Universitätsstraße 15/FE, A-8010 Graz

## Abstract

How will the novel coronavirus evolve? I study a simple epidemiological model, in which mutations may change the properties of the virus and its associated disease stochastically and antigenic drifts allow new variants to partially evade immunity. I show analytically that variants with higher infectiousness, longer disease duration, and shorter latent period prove to be fitter. "Smart" containment policies targeting symptomatic individuals may redirect the evolution of the virus, as they give an edge to variants with a longer incubation period and a higher share of asymptomatic infections. Reduced mortality, on the other hand, does not per se prove to be an evolutionary advantage. I then implement this model as an agent-based simulation model in order to explore its aggregate dynamics. Monte Carlo simulations show that a) containment policy design has an impact on both speed and direction of viral evolution, b) the virus may circulate in the population indefinitely, provided that containment efforts are too relaxed and the propensity of the virus to escape immunity is high enough, and crucially c) that it may not be possible to distinguish between a slowly and a rapidly evolving virus by looking only at short-term epidemiological outcomes. Thus, what looks like a successful mitigation strategy in the short run, may prove to have devastating long-run effects. These results suggest that optimal containment policy must take the propensity of the virus to mutate and escape immunity into account, strengthening the case for genetic and antigenic surveillance even in the early stages of an epidemic.

# Endogenous viral mutations, evolutionary selection, and containment policy design

Patrick Mellacher[1]

[1]University of Graz, Graz Schumpeter Centre

patrick.mellacher@uni-graz.at

First version: 09th of July 2021

This version: 13th of December 2021

**Abstract**: How will the novel coronavirus evolve? I study a simple epidemiological model, in which mutations may change the properties of the virus and its associated disease stochastically and antigenic drifts allow new variants to partially evade immunity. I show analytically that variants with higher infectiousness, longer disease duration, and shorter latent period prove to be fitter. "Smart" containment policies targeting symptomatic individuals may redirect the evolution of the virus, as they give an edge to variants with a longer incubation period and a higher share of asymptomatic infections. Reduced mortality, on the other hand, does not per se prove to be an evolutionary advantage. I then implement this model as an agent-based simulation model in order to explore its aggregate dynamics. Monte Carlo simulations show that a) containment policy design has an impact on both speed and direction of viral evolution, b) the virus may circulate in the population indefinitely, provided that containment efforts are too relaxed and the propensity of the virus to escape immunity is high enough, and crucially c) that it may not be possible to distinguish between a slowly and a rapidly evolving virus by looking only at short-term epidemiological outcomes. Thus, what looks like a successful mitigation strategy in the short run, may prove to have devastating long-run effects. These results suggest that optimal containment policy must take the propensity of the virus to mutate and escape immunity into account, strengthening the case for genetic and antigenic surveillance even in the early stages of an epidemic.

**JEL-Codes**: C63, H12, I10, I18

**Funding**: Open Access Funding provided by the University of Graz.

**Conflicts of interest**: None.

**Code Availability**: The Code of the simulation model and a graphical user interface is available at https://github.com/patrickmellacher/viralmutations.

# 1 Introduction

In the autumn of 2021 it has become clear, that even wide-spread access to vaccines has not (yet) eliminated the threat of Covid-19, as new variants prove to be fitter and better able to escape immunity (Bernal et al. 2021; Hoffmann et al. 2021a; Hoffmann et al. 2021b; Wall et al. 2021). Mutations have already started to become a concern in mid-2020 (Korber et al. 2020), whereas other early studies concluded that the evolutionary pace of SARS-CoV2 was too low to endanger vaccine efficacy (Dearlove et al. 2020). In the beginning of 2021, however, most experts believed that Covid-19 will become "endemic", i.e. circulate perpetually in varied forms at least in certain areas (e.g. countries with low access to vaccines) (Phillips 2021).

In this article, I aim to shed light on two research questions which naturally arise in this situation: a) in which direction will the SARS-CoV2 virus causing Covid-19 evolve?, and b) which aggregate dynamics can we expect from an evolving virus in terms of infections and fatalities? In order to study these topics, I develop a parsimonious epidemiological model that simultaneously captures two types of viral evolution:

a) Genetic variation: Mutations can change the characteristics of a virus in a way that alter its evolutionary fitness, e.g. by increasing its transmissibility (Chen et al. 2020).

b) Antigenic variation (Yuan et al. 2021): Antigenic drift is a process in which the virus changes its antigenic profile. Since antibody response is targeted at the antigenic profile, an antigenic drift allows a virus to escape (some) immunity previously acquired by the viral host. Research on the influenza virus show that genetic change is more gradual than antigenic evolution and that immune escape depends on the antigenic distance between two variants (Smith et al. 2004).

Both types of variation are subject to natural selection, which favors "useful" (Darwin 1859) variation, i.e. variation which increases the growth rate of the number of people infected by a variant. A recent study suggested that the viral evolution of SARS-CoV2 has been accelerating (McCarthy et al. 2021).

Theoretical modelling of fitness evolution usually relies on the so-called fitness landscape approach first introduced by Wright (1932). In this model, mutations cause a species (e.g. virus) to move along an n-dimensional landscape. Each spot in this landscape represents a phenotype associated with a given fitness value, which may be static or change over time due to, for instance, environmental effects (e.g. Wilke et al. 2001). This landscape may be multi-peaked: Such an approach (using a one-dimensional landscape) was recently used to study viral evolution by e.g. Rüdiger et al. (2020). It may also, however, be single-peaked, a case in which the species continuously approaches the peak. Gurevich et al. (2021) study the evolutionary competition between two distinct strains and find that increased testing favors a test-evasive strain.

In order to capture partial strain-dependent immunity, Roche et al. (2011) develop an agent-based model where previous infections provide partial cross-immunity depending on the evolutionary distance between the infecting variant and the variant that had caused an infection in the past (as observed empirically by Smith et al. 2004). An alternative approach is chosen by Griffin et al. (2020), who develop a parsimonious model with strain-dependent immunity that does not need to store the "antigenic history" (ibid) of each agent.

This paper contributes to the literature on the theoretical modelling (the impact) of viral mutations and – more specifically – Covid-19 variants (e.g. Roche et al. 2011; Basurto et al. 2021; Buckee et al. 2007; Cao et al. 2021; Gabler et al. 2021; Gurevich et al. 2021; Gordo et al. 2009; Griffin et al. 2020; Halley et al. 2021; Marquioni and Aguiar 2021; Pageaud et al. 2021; Rella et al. 2021; Rüdiger et al. 2020; Williams et al. 2021). My contribution is to a) introduce a parsimonious model of endogenous viral evolution capturing both genetic and antigenic variation (i.e. evolving intrinsic and extrinsic fitness, see Smith et al. 2004), as well as imperfect cross-immunity, and b) use this framework to study aggregate dynamics and the direction of viral evolution under varying containment policies.

This paper also contributes to the literature using agent-based modelling to studying the Covid-19 crisis (e.g. Basurto et al. 2021; Delli Gatti and Reissl 2020; Dignum et al. 2020; Gabler et al. 2021; Kerr et al. 2021; Lasser et al. 2020; Mellacher 2020, 2021a; Silva et al. 2020; Vermeulen et al. 2021; Wallentin et al. 2020).

Finally, this paper is also related to a literature to the study of evolutionary processes in economics as pioneered by Nelson and Winter (1982). This approach has been adopted in a particularly fruitful way using agent-based models in the field of innovation economics (e.g. Ma and Nakamori 2004; Dosi et al. 2010), which *inter alia* considers the case of directed change due to evolutionary selection (Fanti 2020; Hötte 2020; Mellacher and Scheuer 2021). Modelling the behavior of agents as an adaptive (boundedly rational) processes is another highly promising field of study where evolutionary processes are employed by economists. A recent example for such a model is developed by Lux (2021), who extends the classical SIR model to include adaptive endogenous social distancing.

The rest of this paper is structured as follows: The second section presents the basic model and analyzes the direction of evolution under varying containment scenarios analytically. The third section discusses the implementation of this model as an agent-based simulation model (ABM). The fourth section shows the results of a quantitative analysis of the ABM using Monte Carlo simulations. Finally, section five concludes.

## 2 SEPAIRD model and analytical results

*2.1 Baseline model (without containment policies)*

This section describes the baseline model using differential equations in order to investigate its basic properties analytically. In order to capture pre-symptomatic and asymptomatic infections, I extend the classical SIR model to incorporate the following compartments (which also denote the states to which agents in the ABM may belong): Susceptible (S), exposed (E), pre-symptomatic (P), (permanently) asymptomatic (A), (symptomatic) infected (I), recovered (R) or dead (D). Susceptibles may become infected with the virus, if they meet a person who belongs to the compartments P, A or I. Once infected, a susceptible becomes exposed. This is known as a latent period, in which the person neither displays any symptoms, nor is infectious. After the latent period, a person becomes pre-symptomatic. In this case, agents are able to transmit the disease, but do not display any symptoms (yet). Pre-symptomatic agents may later become either symptomatic infected or stay permanently asymptomatic. Finally, permanently asymptomatic will recover, whereas symptomatic infected may either recover or die. A similar approach is used by Lee et al. (2009). This rather detailed specification is necessary to disentangle the effects of various properties of the virus and its associated disease on the effective reproduction number $R_t$, which governs the growth rate of the number of infected.

Like the standard SIR model, this model operates on the simplifying homogenous mixing assumption, i.e. every member of compartment i has an equal probability to meet a member of compartment j.

Figure 1 gives a graphical overview of the model:

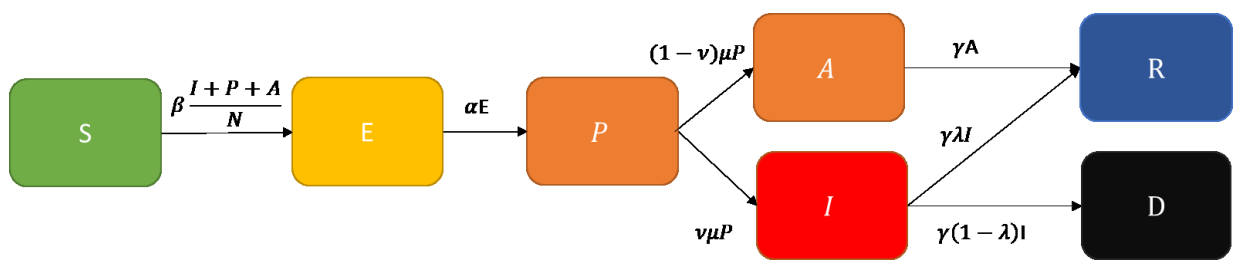

**Figure 1**: Variant-specific SEPAIRD model

The laws of motion (omitting mutations and variants) of this model are given as follows, where $\beta$ denotes the average number of infectious contacts per period, $\frac{1}{\alpha}$ the average latent period, $\frac{1}{\mu}$ the average pre-symptomatic time, $\nu$ the share of symptomatic infections ($0 < \nu < 1$), $\frac{1}{\gamma}$ the average duration of symptoms (if the infection takes a symptomatic course), and $\lambda$ the chance to survive a symptomatic infection ($0 < \lambda < 1$):

$$\dot{S}(t) = -S(t)\beta\left(\frac{I(t) + P(t) + A(t)}{N(t)}\right) \tag{1}$$

$$\dot{E}(t) = S(t)\beta\left(\frac{I(t) + P(t) + A(t)}{N(t)}\right) - E(t)\alpha \tag{2}$$

$$\dot{P}(t) = E(t)\alpha - P(t)\mu \tag{3}$$

$$\dot{A}(t) = P(t)\mu(1-\nu) - A(t)\gamma \tag{4}$$

$$\dot{I}(t) = P(t)\mu\nu - I(t)\gamma \tag{5}$$

$$\dot{R}(t) = A(t)\gamma + I(t)\lambda\gamma \tag{6}$$

$$\dot{D}(t) = I(t)(1-\lambda)\gamma \tag{7}$$

The basic reproduction number is defined as the number of people infected by one infected person in an otherwise susceptible population. Since an infected person may belong to the compartments E (where they are not infectious), P, A or I (where they are infectious), it can be calculated by multiplying the number social contacts per time period with the time spent, i.e.

$$R_0 = \frac{\beta}{\mu} + \frac{\beta(1-\nu)}{\gamma} + \frac{\beta\nu}{\gamma} \tag{8}$$

which collapses to:

$$R_0 = \frac{\beta}{\mu} + \frac{\beta}{\gamma} \tag{9}$$

The effective reproduction number $R_t$ is the number of people infected by one infected person in a population which otherwise does *not* only consist of susceptibles. Assuming that infectious contacts are not directed towards any group (i.e. homogenous mixing), the share of these contacts with susceptibles is given by the number of susceptibles in the population, i.e. $\frac{S(t)}{N(t)}$. In the absence of any containment measures, the effective reproduction number is thus given by:

$$R_t = \left(\frac{\beta}{\mu} + \frac{\beta}{\gamma}\right)\frac{S(t)}{N(t)} \tag{10}$$

If $R_t < 1$, the virus will die out. If $R_t > 1$, the number of infected will grow exponentially. As such, $R_t$ is the key metric to capture the evolutionary fitness of a new variant. Accordingly, we can use partial derivatives to investigate how changes in the viral properties may affect $R_t$, i.e. to investigate the shape of the fitness landscape with regard to each viral property.

Differentiating $R_t$ with respect to each parameter and variable shows that it increases with $\beta$ (i.e. the transmissibility), decreases with $\gamma$ and $\mu$ (i.e. increases with the time in which an individual is infectious $\frac{1}{\mu} + \frac{1}{\gamma}$), increases with $S(t)$ (i.e. the susceptibles) and decreases with $N(t)$ (i.e. the total population).

From this analysis follows easily that those mutations are more successful, which increase the transmissibility and the infectious time. From the relationship with $S(t)$ follows that those mutations are more successful that a) exhibit lower cross-immunity (i.e. circumvents pre-existent immunity against other variants more effectively), and b) that are able to attack more quickly before the recipients are able to obtain cross-immunity from other variants, i.e. have a lower latent period. A decrease in $N(t)$ is achieved by a higher lethality of the disease caused by the variant. Such a decrease, however, affects all variants simultaneously and thus cannot be assumed to provide an evolutionary advantage to any specific variant.

*2.2 Uniform social distancing*

Uniform social distancing (which could also be imagined as a blanket lockdown or compulsory mask wearing) reduces the number of social contacts (or their infectiousness) for all individuals in the same way.[1] This can be modeled by substituting $\beta$ with $\beta(1-\delta)$, where $\delta$ denotes the share of social contacts avoided during each period (or alternatively the reduction of infectiousness of each social contact). Accordingly, the laws of motion regarding S and E change:

$$\dot{S}(t) = -S(t)\beta(1-\delta)\left(\frac{I(t) + P(t) + A(t)}{N(t)}\right) \tag{11}$$

$$\dot{E}(t) = S(t)\beta(1-\delta)\left(\frac{I(t) + P(t) + A(t)}{N(t)}\right) - E(t)\alpha \tag{12}$$

The basic reproduction number is now:

$$R_0 = \left(\frac{\beta}{\mu} + \frac{\beta}{\gamma}\right)(1-\delta) \tag{13}$$

With the new effective reproduction number given by:

$$R_t = \left(\frac{\beta}{\mu} + \frac{\beta}{\gamma}\right)(1-\delta)\frac{S(t)}{N(t)} \tag{14}$$

---

[1] Please note that reducing infectiousness is only a perfect substitute to reducing social contacts in the standard homogeneous mixing SIR-framework. Gutin et al. (2021) develop a network-based SIR model and show that the effect of social distancing depends on the network structure of the social interactions.

This does not change any of the considerations above. If, however, $\delta$ is interpreted as a reduction of infectiousness due to face masks or other protective equipment, any variants that circumvent these measures more effectively also have an evolutionary advantage.

*2.3 Isolation of symptomatic cases:*

Another widespread measure to slow the spread of Covid-19 is to isolate symptomatic cases (or to encourage self-isolation upon developing symptoms). If we assume that all symptomatic cases are isolated, this changes the laws of motion regarding S and E accordingly (i.e. individuals in compartment I do not any longer spread the virus), if we assume that this measure is not combined with uniform social distancing):

$$\dot{S}(t) = -S(t)\beta\left(\frac{P(t) + A(t)}{N(t)}\right) \tag{15}$$

$$\dot{E}(t) = S(t)\beta\left(\frac{P(t) + A(t)}{N(t)}\right) - E(t)\alpha \tag{16}$$

Adapting equation 8, the basic reproduction number is now:

$$R_0 = \frac{\beta}{\mu} + \frac{\beta(1-\nu)}{\gamma} \tag{17}$$

Accordingly, the effective reproduction number is:

$$R_t = \left(\frac{\beta}{\mu} + \frac{\beta(1-\nu)}{\gamma}\right)\frac{S(t)}{N(t)} \tag{18}$$

In this case, the results regarding the evolutionary advantage of an increase in $\beta$, a decrease in $\gamma$ and $\mu$, an increase in $S(t)$ still hold. In addition to that, however, $R_t$ decreases with $\nu$ (i.e. the share of symptomatic infections) and increases with the pre-symptomatic phase $\mu$, even if the total duration of being infectious ($\mu + \gamma$) is constant.[2]

If the lethality depends mechanically on the share of symptomatic infections, as suggested by above formulation (i.e. the chance to survive a symptomatic infection $\lambda$ is unchanged by a change in the share of symptomatic infections), the evolutionary selection mechanism also favors less deadly variants as a

---

[2] Please note that this result holds even if we would assume that asymptomatic individuals are less infectious due to biological reasons (for instance, because coughing individuals spread the virus faster). In order to test this, replace the rate of infectious contacts $\beta$ with $\beta_X$ for asymptomatic and pre-symptomatic individuals and with $\beta_Y$ for symptomatic individuals. If all symptomatic individuals are isolated, the effective reproduction number becomes $R_t = \left(\frac{\beta_X}{\mu} + \frac{\beta_X(1-\nu)}{\gamma}\right)\frac{S(t)}{N(t)}$. This does not affect our results and such a change does thus not result in a trade-off for the direction of viral evolution. If symptomatic individuals are not isolated, however, assuming that $\beta_X < \beta_Y$ would redirect viral evolution towards more symptomatic infections and a shorter incubation period.

*side effect* of favoring asymptomatic infections. If, however, the survival chance is independent from the share of symptomatic infections, lethality is unaffected by evolutionary selection.

**3 Agent-based model**

Due to the interactions between the different variants of the virus, the outcomes of a full model covering endogenous variants cannot be analyzed purely by differential equations. I thus implement this model as a simple open-source agent-based simulation model in NetLogo (Wilensky 1999)[3] to a) confirm the analytical predictions of the model behavior and b) explore aggregate dynamics under varying scenarios.

In order to capture mutations, the interplay between the variants, while at the same time keeping the analysis as simple as possible, I make the following assumptions:

1.) Each agent may only be infected by one virus variant simultaneously.
2.) Whenever an individual is infected, the virus may mutate, creating a new variant which is an offspring of the infecting ("parent") variant.
3.) A mutation randomly changes the properties of the virus (infectiousness, latent period, share of asymptomatic infections, incubation period, disease duration, lethality) with the means given by the actual values of the parent variant (see equation 19).
4.) Each virus variant belongs has an antigenic cluster that may change during a mutation, i.e. a mutation may be coupled with an antigenic drift.[4]
5.) A previous infection within the same antigenic cluster provides perfect cross-immunity between variants. It may provide cross-immunity between antigenic clusters depending on the antigenic (evolutionary) distance between two antigenic clusters.
6.) Each agent has an equal probability to meet another agent (i.e. homogeneous mixing).

The following sequence of events occurs during each simulation step:

1.) Infected agents meet other agents and may infect them, a process in which mutations and antigenic drifts may occur (described in more detail in subsection *3.1*).
2.) The disease progresses, i.e. agents may become infectious, develop symptoms and become isolated (if such a containment policy is active), recover or die (see subsection *3.2*).

---

[3] The model features a Graphical User Interface and can be obtained at https://github.com/patrickmellacher/viralmutations.
[4] In the case of the (well researched) influenza viruses, the genetic evolution is more gradual (and less punctuated) than the antigenic evolution, and influenza variants (strains) thus group into antigenic clusters (See Smith et al. 2004).

3.) Infection statistics are updated (see subsection *3.3*).

*3.1 Infections, mutations and antigenic drifts*

In order to save computational resources, the model only explicitly processes social contacts of infectious agents who are not isolated. This concerns agents who are asymptomatic (A in the notation chosen in section 2 of this paper), pre-symptomatic (P) and – depending on policy – also symptomatic infectious (I). Each agent of these types randomly meets $\eta$ other agents who are alive. Each social contact of an agent infected with variant $m$ with another agent who is neither immune to variant $m$, nor currently infected with any variant, causes an infection with probability $i_m$.[5]

Each infection may cause a mutation with a constant probability $\phi$. In this case, the newly infected agent is the first carrier of the newly emerged variant. During each mutation, all properties of the virus and its associated disease, namely infectiousness, latent period, incubation period, total infection duration, share of asymptomatic infections and lethality, are subject to stochastic multiplicative change, where $h_{i,k}$ is the property $k$ of the variant $i$, which is an offspring of variant $j$.[6]:

$$h_{i,k} = (1 + \omega_{i,k})h_{j,k} \qquad (19)$$

Where $\omega_{k,i}$ ($\omega_{k,i}$ > - 1) is drawn from a normal distribution with a mean of $\theta$ and standard deviation of $\sigma^I$.[7]

$$\omega_{k,i} \sim N(\theta, \sigma^I) \qquad (20)$$

The mean ($\theta$) is set to 0 in order not to presuppose any direction of evolution. Instead, all properties change stochastically and those changes, which prove to be advantageous, assert themselves endogenously in the competition against the other variants. Thus, "each slight variation, if useful, is preserved" Darwin (1859, 61).

A fraction $\kappa$ of mutations is coupled by an antigenic drift, i.e. able to evade some immunity. Following empirical research on the influenza virus (Smith et al. 2004), this is modeled by assuming that each variant belongs to an "antigenic cluster". Whenever an antigenic drift (i.e. a new immune escaping mutation) occurs, every agent who is immune to its ancestor may become immune to the new antigenic cluster with the probability $\psi^I$.

---

[5] Please note that it is not necessary to disentangle these two effects in a classical SIR-type model as, on average, the following condition holds $\eta\varphi = \beta$, where $\beta$ denotes the number of infectious contacts per period as defined in section 2.

[6] This approach follows a simple multiplicative approach (e.g. Miller et al. 2018) for each dimension determining the fitness separately, as the fitness landscape is single-peaked or flat in each dimension.

[7] In the model code, $\omega_{k,i}$ is set to be -0.99 at minimum in order to avoid any negative values for $h_{i,k}$ even for extreme parameters of the random distribution.

Figure 2 shows an example "phylogenetic tree", which plots the evolution of the virus into different variants. The arrows point from ancestor variants to their descendants. Each descendant differs slightly from its ancestor. Two antigenic drifts (at $V_{1.2}$ and $V_{1.1.2}$) created three distinct antigenic clusters which are colored differently in this figure in order to highlight them. An infection with a variant from the yellow antigenic cluster (e.g. $V_{1.1.2.1}$) provides perfect immunity against other variants belonging to the yellow antigenic cluster. The chance that it also provides cross-immunity against variants belonging to the white antigenic cluster is $\psi^I$, which in turn provides cross-immunity against the green cluster again with probability $\psi^I$.

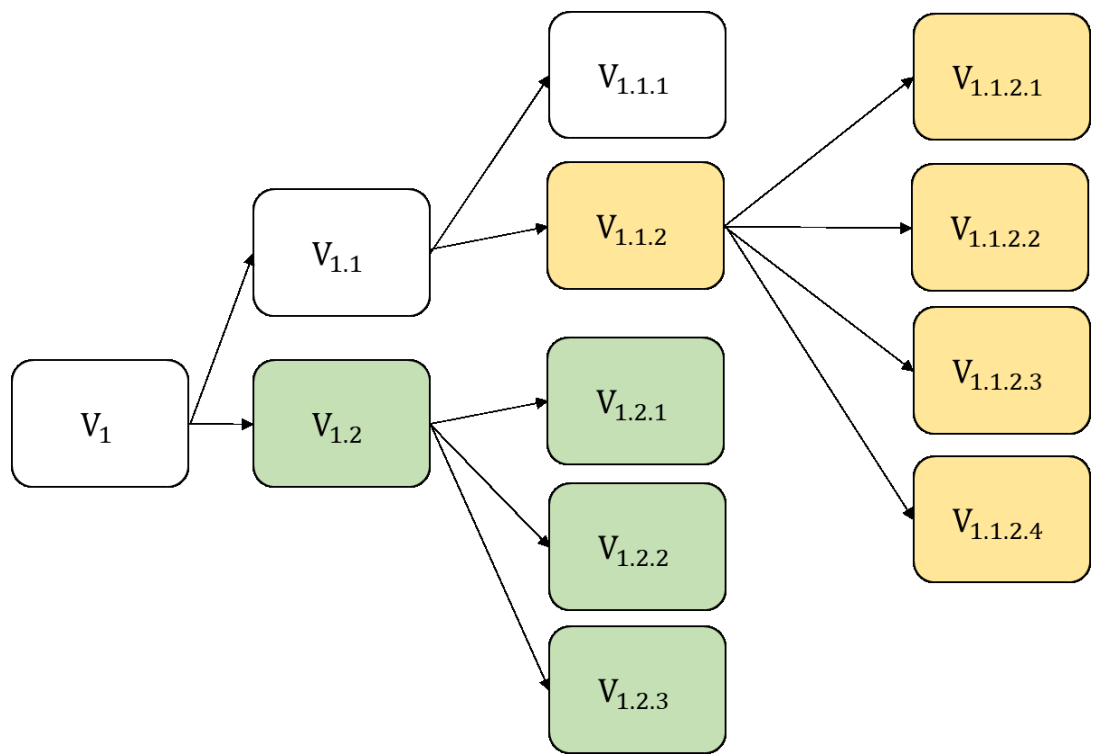

**Figure 2**: Example phylogenetic tree of different variants. Different colors denote different antigenic clusters.

*3.2 Disease progression*

Whenever an individual n is infected with virus m, the actual latent period $l_{n,m}$, incubation time $b_{n,m}$, and the disease duration $d_{n,m}$ (all ≥ 0) are drawn from a normal distribution with the means given by the viral attributes (latent period $l_m$, incubation time $b_m$, disease duration $d_m$) and a standard deviation which is a fraction $\sigma^{II}$ of the respective attribute:

$$l_{n,m} \sim N(l_m, l_m \sigma^{II}) \tag{21}$$

$$b_{n,m} \sim N(b_m, b_m \sigma^{II}) \tag{22}$$

$$d_{n,m} \sim N(d_m, d_m \sigma^{II}) \tag{22}$$

These values are then rounded to the next integer. During each simulation step, a counter (which is initialized with 0 for newly infected individuals) records the time passed since becoming infected. Once it reaches $l_{n,m}$, the individual is infectious and can infect others. At $b_{n,m}$, it develops symptoms with probability $v_m$ and may isolate itself. At $d_m$ it either dies with a probability given by the lethality rate $f_m$ or otherwise recovers and becomes immune against this antigenic cluster.[8] If an agent has acquired immunity against at least one other variant, it additionally benefits from a "cross protection" $\psi^{II}$ that aims to account for the fact that immunity against other strains (or a vaccination) reduce the lethality even if it cannot prevent an infection. In such a case, the probability of dying is given by $f_m(1 - \psi^{II})$.

Using a recursive function, the agent may also acquire immunity against the ancestor and/or descendants of this antigenic cluster with probability $\psi^I$, as well as the ancestor and possible descendants of those antigenic clusters, to which the agent just acquired cross-immunity with the same probability et cetera.

*3.3 Parameters*

The simulation is initialized with the parameters described in Table 1.

**Table 1: Parameters of the simulation**

| Parameter | Symbol | Value |
|---|---|---|
| human agents | $n$ | 10000 |
| initially infected agents | $n_{inf}$ | 10 |
| daily contacts | $\eta$ | 10 |
| initial infectiousness | $i_0$ | 6.25% |
| initial end of the latent period | $l_0$ | 4 |
| initial end of the incubation period | $b_0$ | 6 |
| initial duration | $d_0$ | 8 |
| initial fatality rate | $f_0$ | 1% |
| initial symptomatic chance | $v_0$ | 70% |
| actual time standard deviation | $\sigma^{II}$ | 0.1 |
| probability of a mutation | $\phi$ | 0/0.5/1/2% |
| mutation mean | $\theta$ | 0 |

---

[8] Please note that under this specification, the lethality rate does *not* depend on the probability of developing symptoms.

| mutation standard deviation | $\sigma^I$ | 0.05 |
| cross immunity between antigenic clusters | $\psi^I$ | 0/50/90% |
| cross protection against a lethal infection | $\psi^{II}$ | 90/99% |
| probability of antigenic drift during a mutation | $\kappa$ | 10% |
| Isolation upon developing symptoms | | yes / no |
| social distancing | $\delta$ | 0-80% |

This calibration allows for a basic reproduction number of the SARS-CoV2 wild type of 2.5 and a share of pre-symptomatic infections of 50% as estimated by the CDC (2021). The incubation time, as well as the share of symptomatic cases are also taken from CDC (2021). The cross protection against a lethal infection was derived from Abu-Raddad et al. (2021), who estimate the effectiveness of a vaccine against lethal Covid-19 to be 97.5%. All other parameters are set to replicate stylized facts of the virus, such as a low propensity of the virus to mutate (which is certainly higher in my model than in the real world, as my model is only populated by 10,000 agents), and an even lower propensity for an antigenic drift to occur. Thus, the simulation results should not be interpreted as an accurate quantitative prediction of what will happen, but as an explorative scenario analysis.

**4 Simulation results**

In order to analyze the properties of the model, I rely on Monte Carlo simulations. Specifically, I run each scenario 100 times with fixed random seeds for 500 periods. I then analyze the results using the programming language R (R Core Team 2018) with the ggplot2 package (Wickham 2016). In doing so, I want to a) get an idea of the "mean" simulation result, but also b) about the distribution of results and their statistical significance. I thus rely on quantile regressions to capture aggregate dynamics, and notched box plots to interpret cumulative outputs at a single simulation step.

*4.1 Viral evolution*

In this subsection, I give an overview of viral evolution under varying mutation parameters, and show how they are affected by social distancing. Detailed results concerning the evolution of each property are in line with the analytical predictions and presented in the appendix.

Figures 3-6 show that:

a) Evolutionary selection causes the virus to evolve differently in the face of "smart" containment policies aimed at isolating symptomatic individuals.

b) Social distancing is generally able to curb viral evolution.

c) Contrary to b), levels of social distancing that bring the effective reproduction number of the wild type in a population inhabited by a very large share of susceptibles close to 1 can cause variants to evolve to higher fitness than lower levels of social distancing, if there is high cross-immunity between the antigenic clusters, as can be seen in the bottom right part of the figures.

In order to present these results concisely, I compute mean $R_{0,m}$ and $R_{0,m}^{adapted}$, which accounts for the isolation of infected, in the following way for the active strains $m$ at time step 500[9], where, following the notation introduced above, $\eta$ denotes the number of daily contacts, $i_m$ the infectiousness, $d_m$ the duration, $l_m$ the latent period, $v_m$ the probability of developing symptoms, $b_m$ the incubation time. Please note that the latent period has to be deducted from the disease duration, as $d_m$ covers both the infectious and the pre-infectious ("latent") period:

$$R_{0,m} = \eta i_m (d_m - l_m) \tag{23}$$

$$R_{0,m}^{adapted} = (1 - v_m)\eta i_m (d_m - l_m) + v_m \eta i_m (b_m - l_m) \tag{24}$$

Figures 3 and 4 shows the mean $R_0$ for scenarios in which symptomatic individuals are not isolated or isolated respectively. We can see that isolating individuals exhibiting symptoms causes the virus to die out in more scenarios and to be less fit in the others with regard to $R_0$.

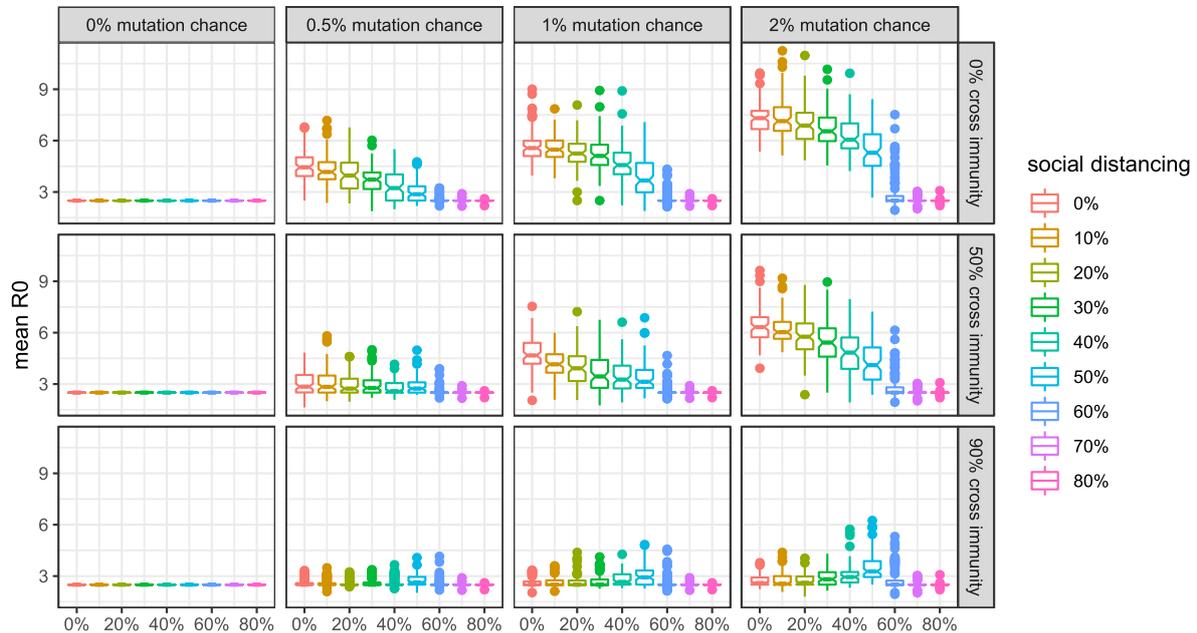

**Figure 3**: Mean R0 of active (or, in case of extinction, last surviving) variants at simulation step 500 <u>without isolation</u> of symptomatic individuals and with 99% cross protection against a lethal infection (notched box plot).

---

[9] If the virus died out before time step 500, it computes the last surviving strain.

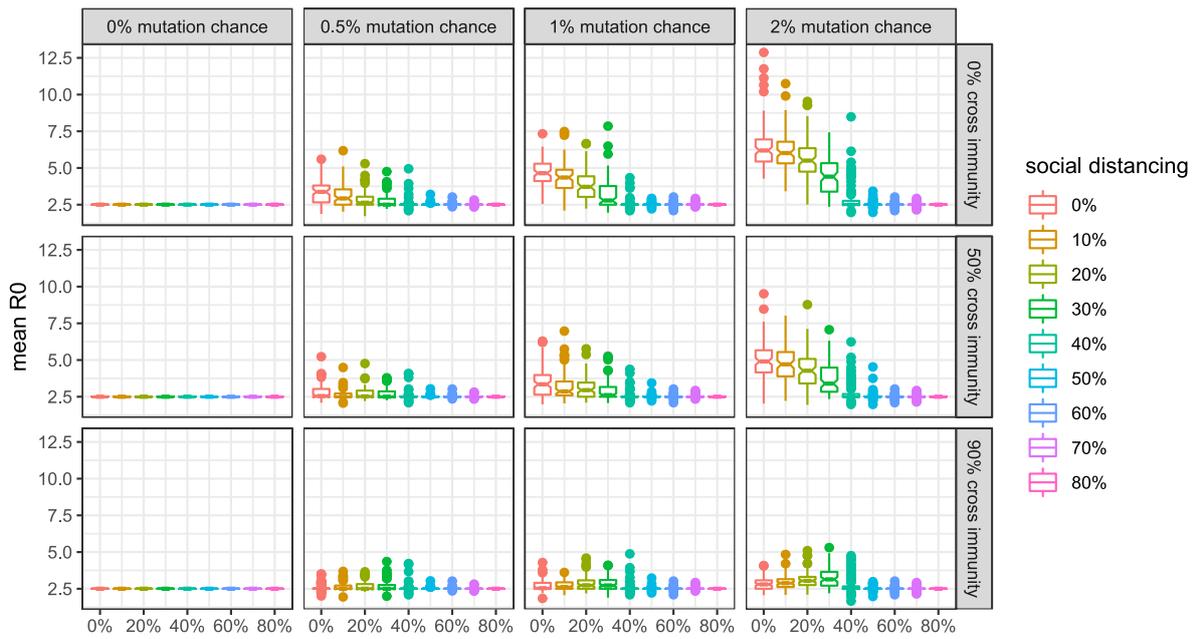

**Figure 4**: Mean R0 of active (or, in case of extinction, last surviving) variants at simulation step 500 <u>with isolation</u> of symptomatic individuals and with 99% cross protection against a lethal infection (notched box plot).

Figures 5 and 6 show the evolution of the mean *relative* adapted $R_0^{adapted}$, i.e. $\frac{R_0^{adapted}}{R_0}$ which describes how efficient isolation policies are in curbing the spread of the virus. If the *relative* adapted $R_0^{adapted}$ is equal to one, isolation of symptomatic individuals does not curb the spread of the virus at all. While isolation policies tend to become more efficient if they are not enacted (see figure 5), adaptive evolution causes them to become less efficient, if they are enacted (see figure 6).

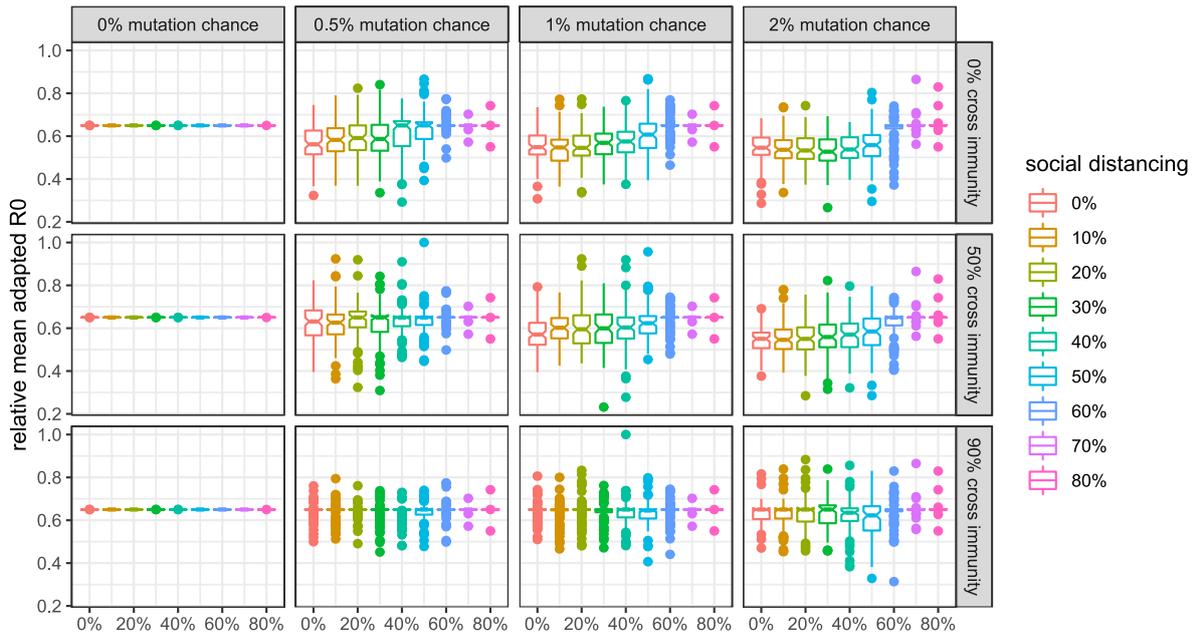

**Figure 5**: Mean $\frac{R_0^{adapted}}{R_0}$ of active (or, in case of extinction, last surviving) variants at simulation step 500 <u>without isolation</u> of symptomatic individuals and with 99% cross protection against a lethal infection (notched box plot).

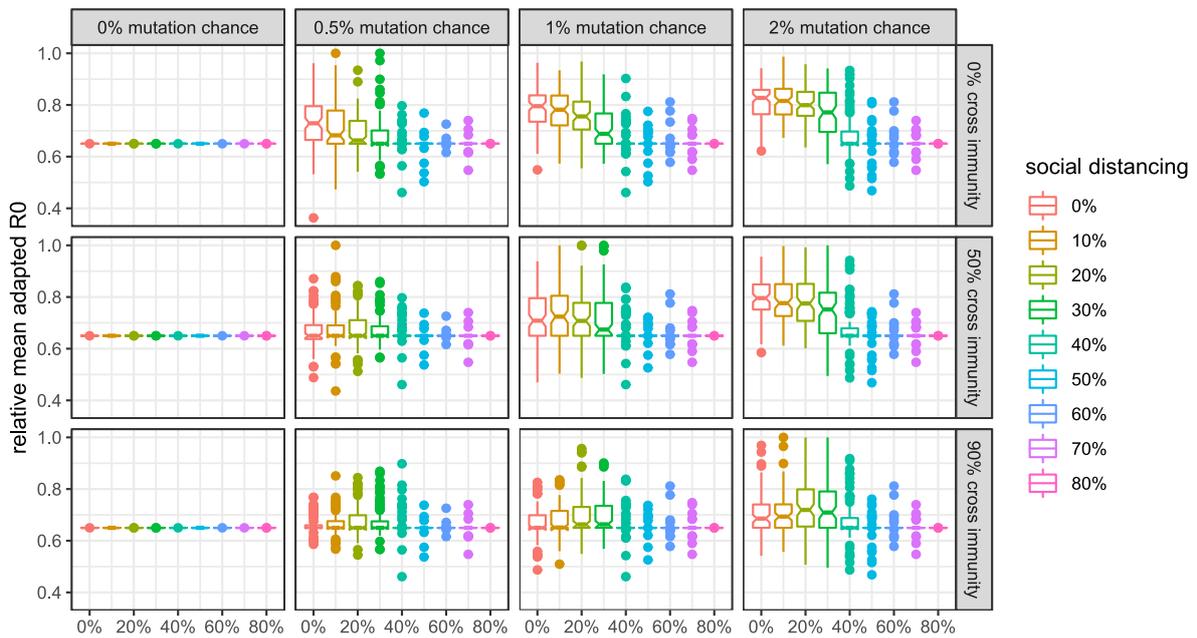

**Figure 6**: Mean $\frac{R_0^{adapted}}{R_0}$ of active (or, in case of extinction, last surviving) variants at simulation step 500 <u>with isolation</u> of symptomatic individuals and with 99% cross protection against a lethal infection (notched box plot).

*4.2 The public health impact of an evolving virus*

This subsection discusses the public health outcomes of an evolving virus and to which extent they can be used to identify the propensity of the virus to mutate. Figures 7 and 8 show public health outcomes in the first 100 time steps of the simulation, which cover (most of) the first wave for almost all scenarios.

There is no visible difference in mortality between the scenarios (assuming a 99% cross protection against a lethal infection) for a given level of social distancing. The number of infected agents is more sensitive to changes, as the second wave of infection is visible for scenarios without any cross-immunity between antigenic clusters and very low levels of social distancing. In more realistic scenarios, however, we also cannot distinguish between the scenarios with regard to the mutation chance and the cross immunity between antigenic clusters by looking only at public health outcomes of the first 100 days.

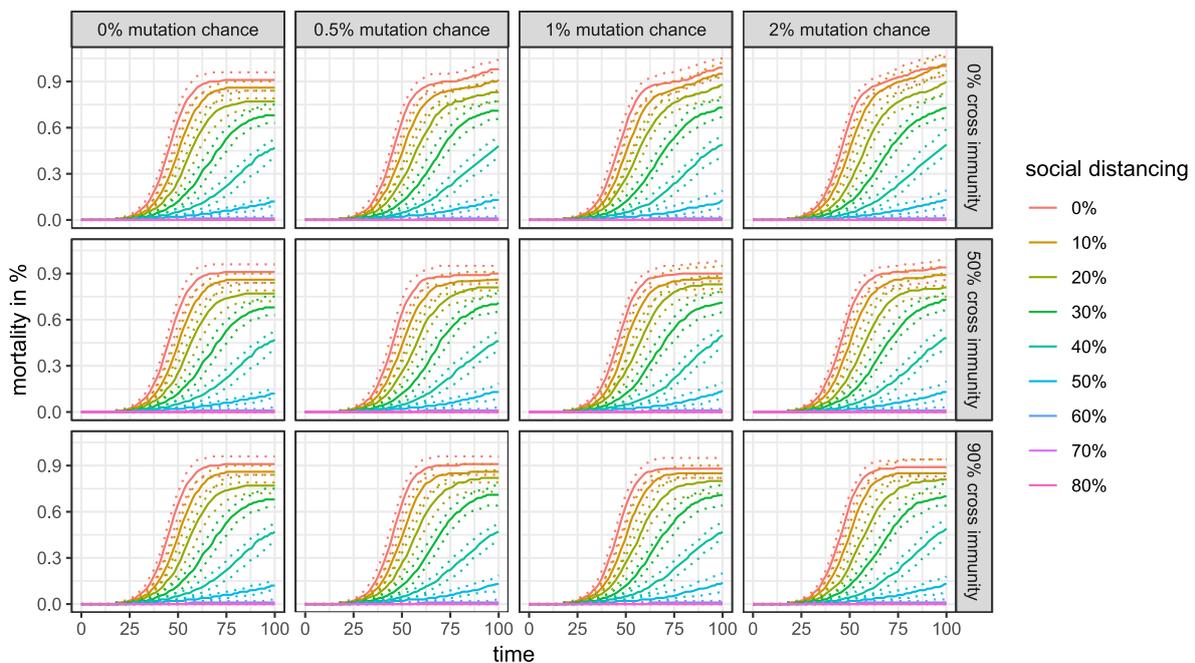

**Figure 7**: Mortality rate in the first 100 simulation steps with 99% cross protection against a lethal infection (quantile regressions)

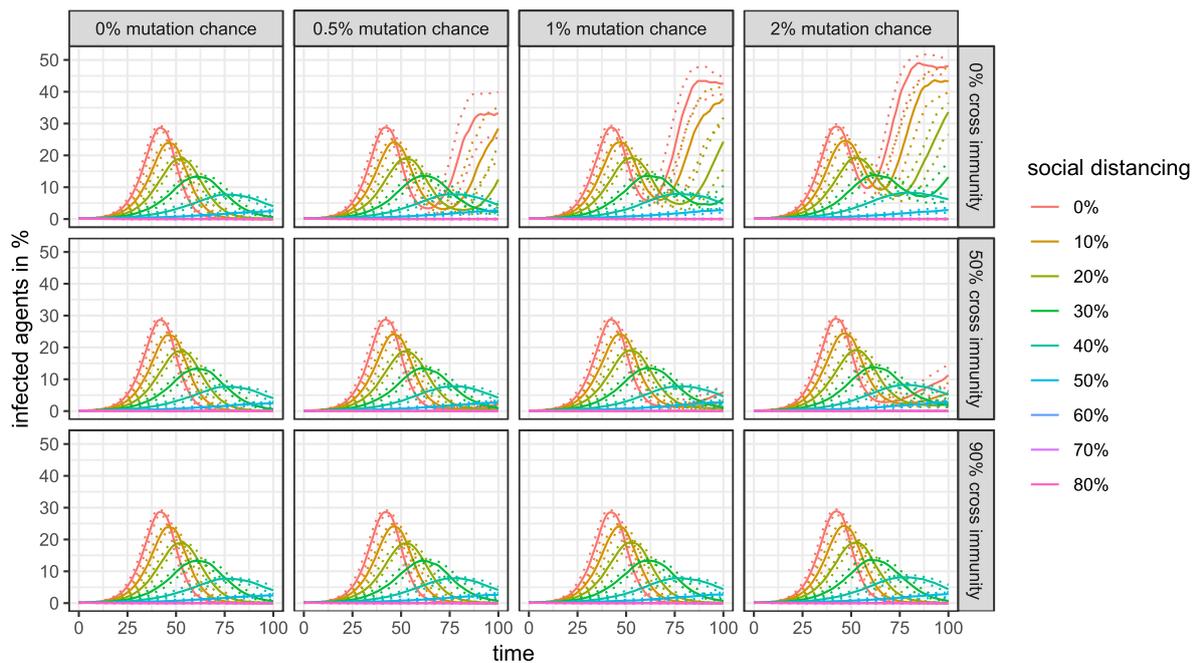

**Figure 8**: Share of infected agents in the first 100 simulation steps with 99% cross protection against a lethal infection (quantile regressions)

Things are different, however, if we look at a longer time horizon. Figure 9 shows the share of infected agents for the first 500 simulation steps, and figures 10 and 11 shows the mortality rate for the same time horizon and cross-protection levels of 99% and 90% respectively. Depending on the levels of cross-protection, the mortality rate added in the "endemic" phase can even surpass the mortality during the first wave.

Moreover, the marginal burden associated with not preventing a full-scale outbreak (i.e. the difference between 50 and 60% social distancing in our scenarios) may drastically increase even in a scenario with high cross-protection and moderate cross-immunity, as mutations and antigenic drifts increase the chances of each individual to become infected at least once in their lifetime.

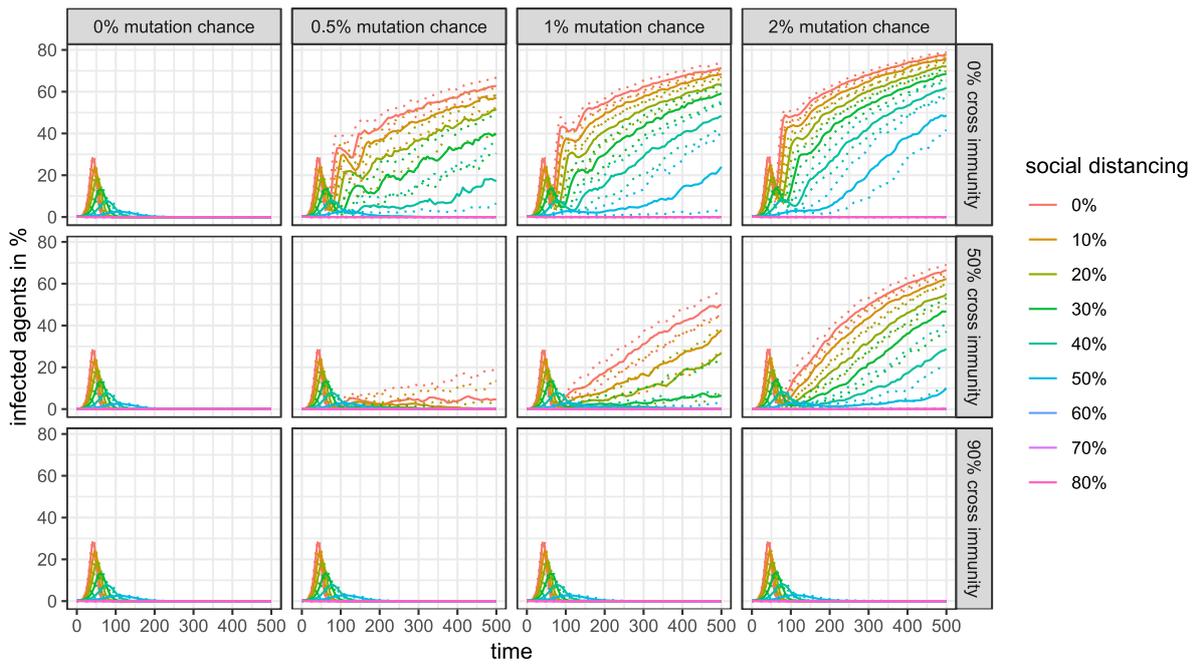

**Figure 9**: Share of infected agents in the first 500 simulation steps with 99% cross protection against a lethal infection (quantile regressions)

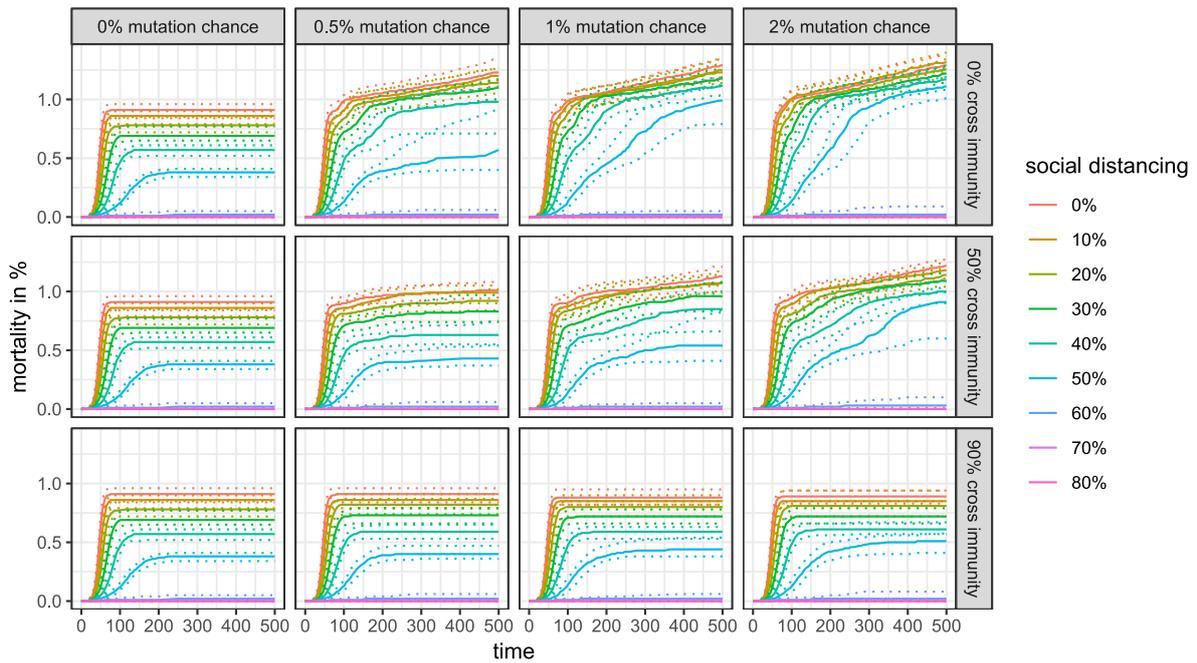

**Figure 10**: Mortality rate in the first 500 simulation steps with 99% cross protection against a lethal infection (quantile regressions)

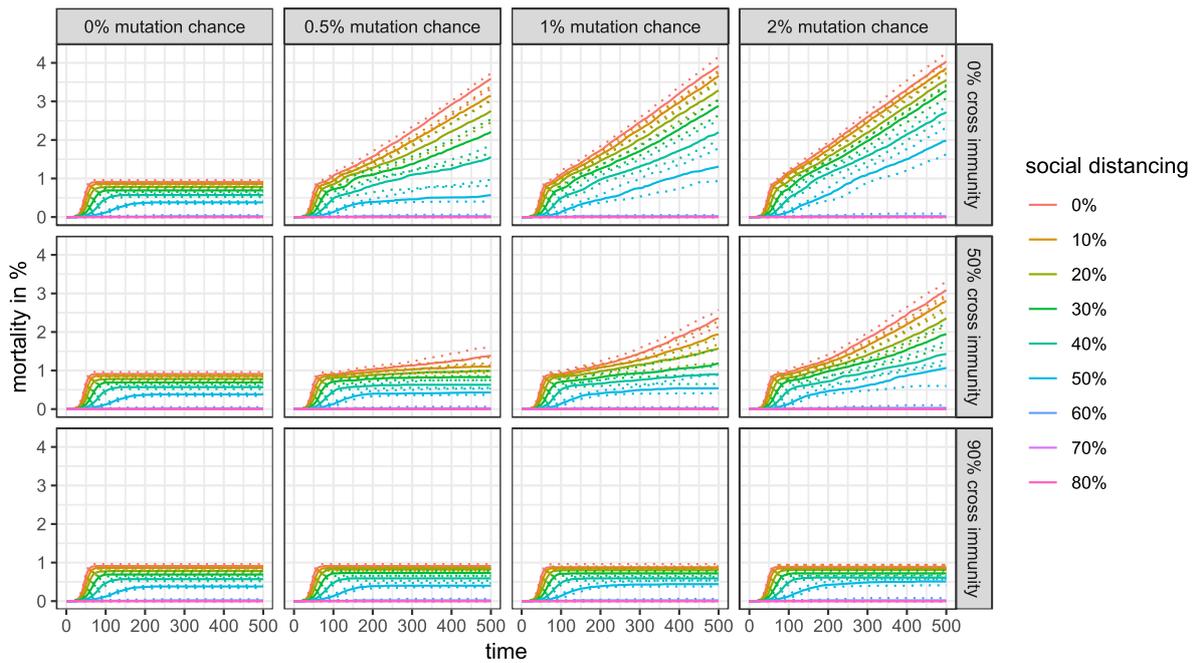

**Figure 11**: Mortality rate in the first 500 simulation steps with 90% cross protection against a lethal infection (quantile regressions)

*4.3 Genetic and antigenic variation*

Figures 12 and 13 show that the genetic and antigenic evolution of the virus are closely connected in my model: figure 8 shows the evolution of the maximum antigenic distance to the wild-type, where two antigenic clusters with an antigenic distance of 1 are separated by only one antigenic drift. Figure 9 shows the mean phylogenetic distance of all active variants to the wild-type, where two variants have a phylogenetic distance of 1 if they are separated by only one mutation.

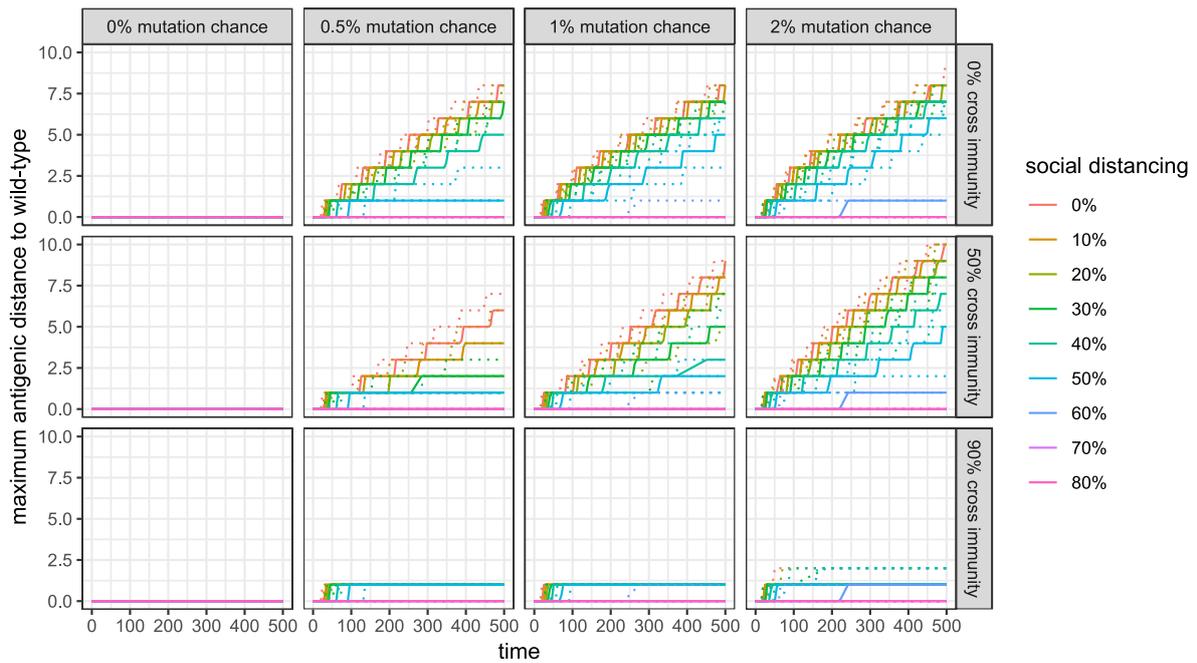

**Figure 12**: Maximum antigenic distance to wild-type in the first 500 simulation steps with 99% cross protection against a lethal infection (quantile regressions)

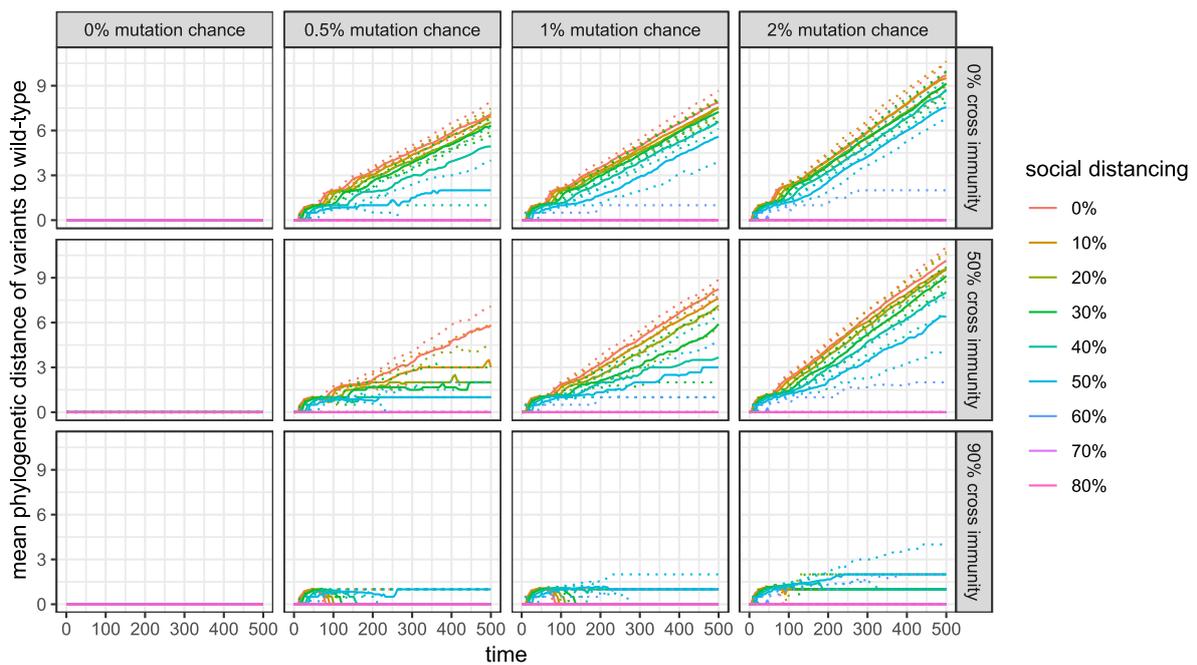

**Figure 13**: Mean phylogenetic distance of active variants to wild-type in the first 500 simulation steps with 99% cross protection against a lethal infection (quantile regressions)

## 5 Conclusion

I developed a simple theoretical model to study the genetic and antigenic evolution of a virus under varying containment scenarios in a stylized way. The properties of the wild-type (i.e. the first variant of the virus prior to any mutations) are calibrated to resemble (infections with) the SARS-CoV2 virus causing Covid-19. Despite its limitations, some of which I outline below, I derived several crucial insights from my model that are empirically testable:

First, containment policies have an impact on the speed of the evolution of the virus. All containment policies that successfully curb the number of infections reduce the number of mutations and thus also the rate at which the virus increases its fitness. If cross-immunity is high, however, the ultimate fitness of a virus may be higher if it circulates in a population engaging in medium-level social distancing than in a population not engaging in social distancing at all due to a slower, but longer, evolution of the virus.

Second, containment policies may also affect the direction of viral evolution. Namely, if symptomatic individuals are isolated, viral fitness increases with the incubation time and the share of asymptomatic infections. Those traits then assert themselves via the endogenous process of evolutionary selection, thus making "smart" isolation policies less effective over time.

Third, it is often not possible to distinguish between an "endemic" scenario, in which variations of the virus persist by continuously evolving and escaping immunity and a non-endemic scenario by looking only at public health outcomes during the first wave of infections. What seems to be a successful "herd immunity" strategy may plant the seeds for a long-term presence of a potentially lethal virus. Thus, it is crucial to monitor the number and severity of reinfections already in the early phases of an epidemic in order to assess a virus' potential to become endemic and optimally design containment policies accordingly.

My model is limited in various ways: First, it assumes that each property of the virus and associated disease may be subject to the same process of stochastic change, which is an obvious stylization. Second, it assumes a lethality rate, and, more generally, all properties of the disease which are uniform across the population. Relaxing this assumption in order to account for e.g. an age-dependent severity of the disease, may influence the results by reducing mortality even in case of lower levels of cross-protection gained from a past infection (or vaccination). Third, my model does not consider behavioral heterogeneity within the population (Mellacher 2021b) or endogenous changes in the social distancing behavior of the population (Lux 2021; Proaño and Makarewicz 2021). Fourth, my model purely concentrates on public health outcomes and thus does not consider any societal or economic impact of social distancing.

Further research on this topic could go in several directions: First, it could address the limitations of this model by including endogenous viral evolution into a larger economic-epidemiological agent-based model (e.g. Basurto et al. 2020; Delli Gatti and Reissl 2020; Mellacher 2020). Second, one could aim to calibrate the parameters governing the viral evolution with empirical data in order to provide more accurate empirical forecasts. Third, this simple approach could be extended to incorporate crucial aspects of viral evolution that are not yet covered in these large agent-based models, but seem to play an important role empirically, such as vaccination campaigns or the spread of the virus in a multi-country world.

**Acknowledgements:** I thank two anonymous referees and the editor, Prof. Thomas Lux, for their valuable and timely feedback, as well as their helpful suggestions for improvement. I also thank the participants of the 28[th] PhD/PostDoc ABM Webinar and the COLIBRI Day 2021 for their comments. All errors are mine.

**7 Appendix:**

This appendix shows the mean viral properties of surviving variants at simulation step 500 (or of those variants which died out last). Unless stated explicitly, these simulations cover the results of simulations without isolation of symptomatic individuals. These simulation results are in line with the analytical predictions presented in section 2.

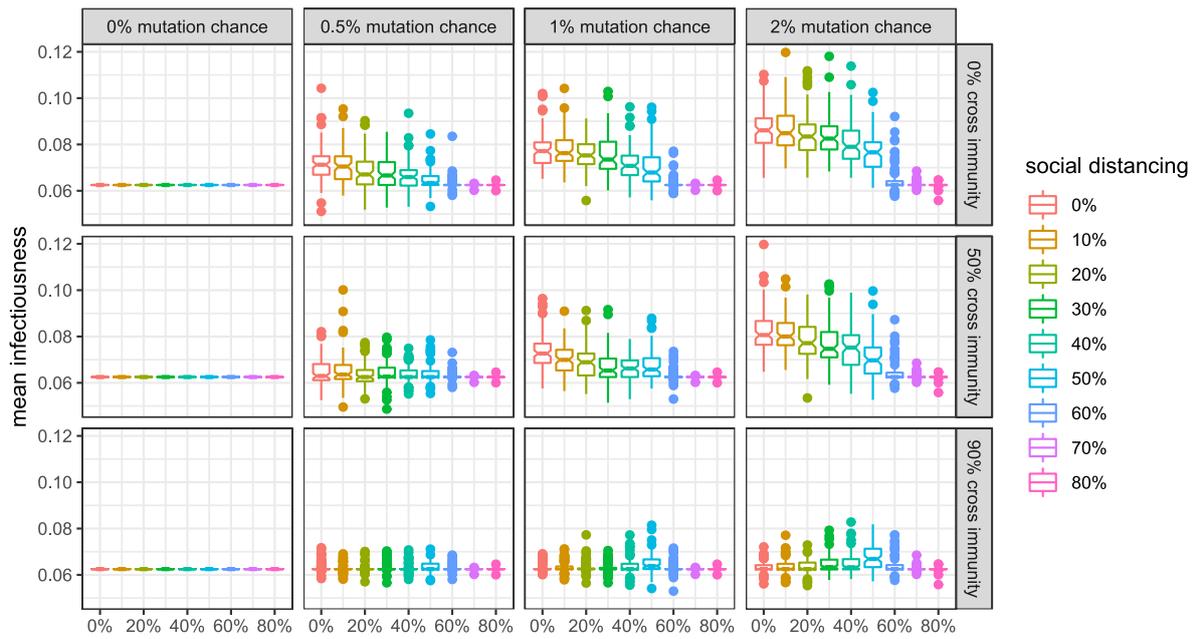

**Figure 14**: Mean infectiousness of active (or, in case of extinction, last surviving) variants at simulation step 500 with 99% cross protection against a lethal infection (notched box plot).

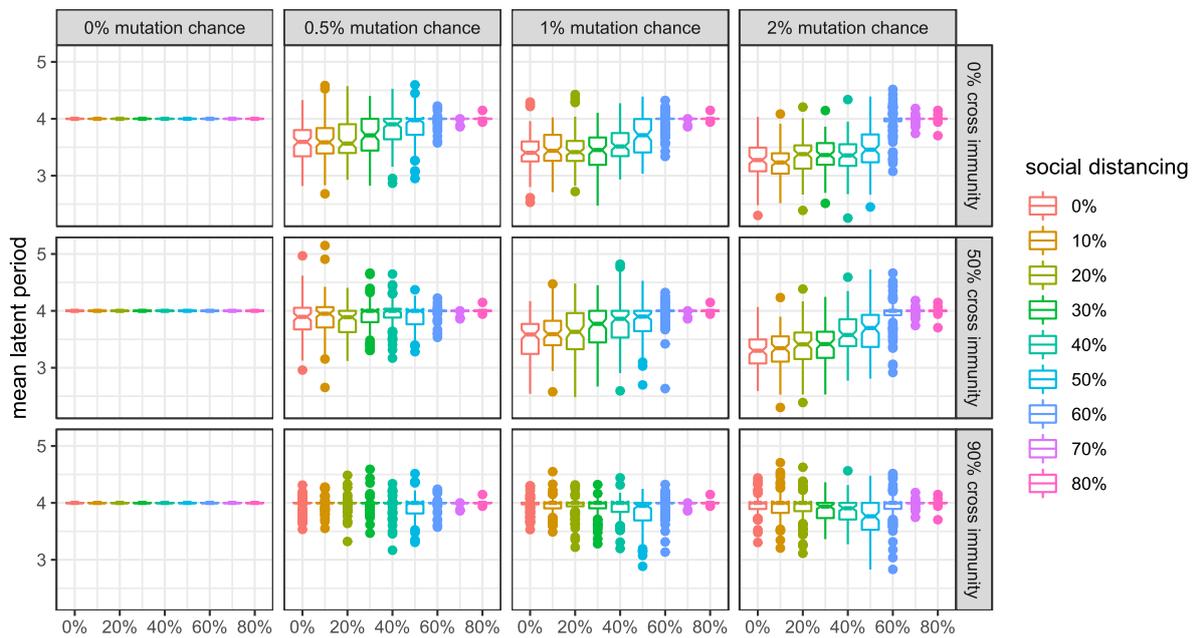

**Figure 15**: Mean latent period of active (or, in case of extinction, last surviving) variants at simulation step 500 with 99% cross protection against a lethal infection (notched box plot).

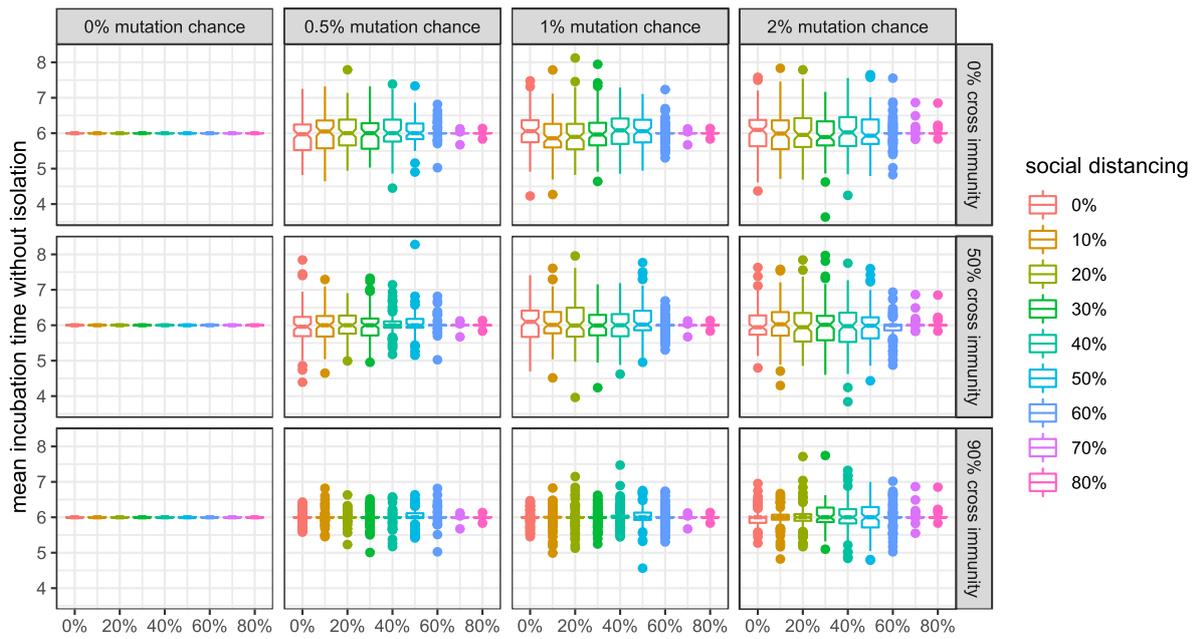

**Figure 16**: Mean incubation period of active (or, in case of extinction, last surviving) variants at simulation step 500 <u>without isolation of symptomatic individuals</u> and with 99% cross protection against a lethal infection (notched box plot).

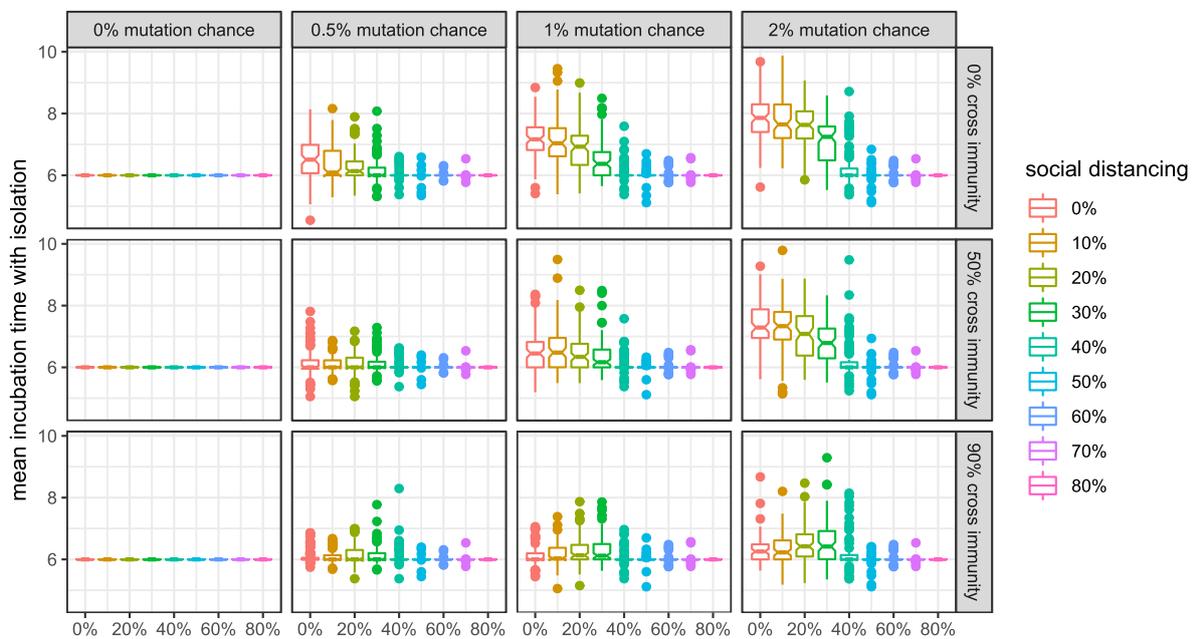

**Figure 17**: Mean incubation period of active (or, in case of extinction, last surviving) variants at simulation step 500 <u>with isolation of symptomatic individuals</u> and with 99% cross protection against a lethal infection (notched box plot).

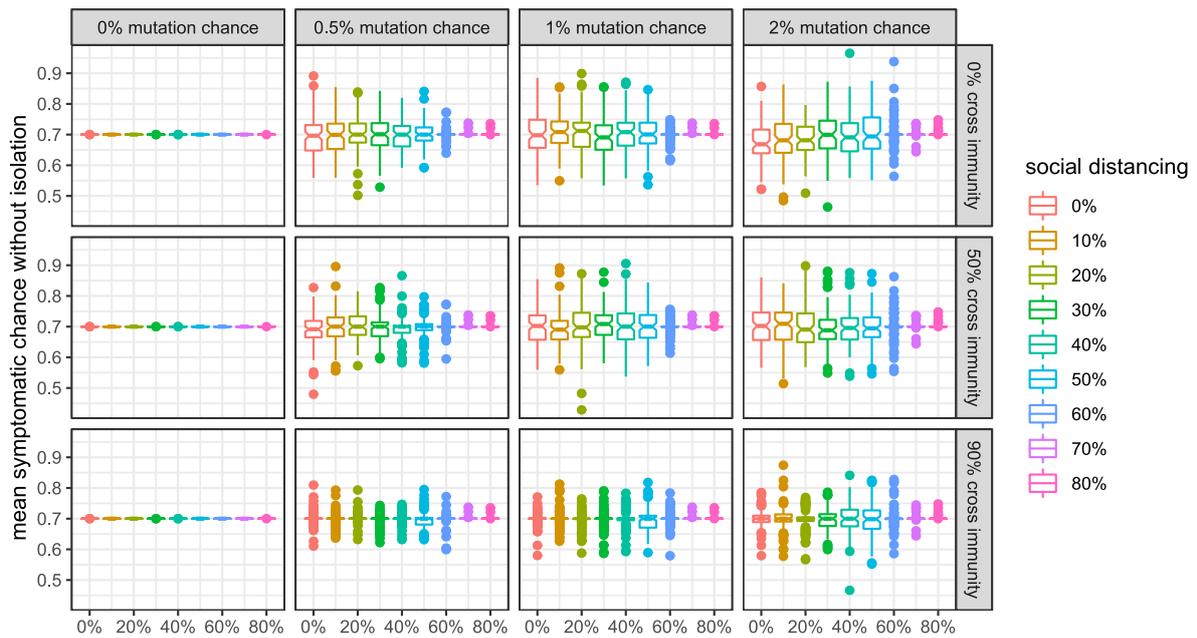

**Figure 18**: Mean probability of a symptomatic course of active (or, in case of extinction, last surviving) variants at simulation step 500 <u>without isolation of symptomatic individuals</u> and with 99% cross protection against a lethal infection (notched box plot).

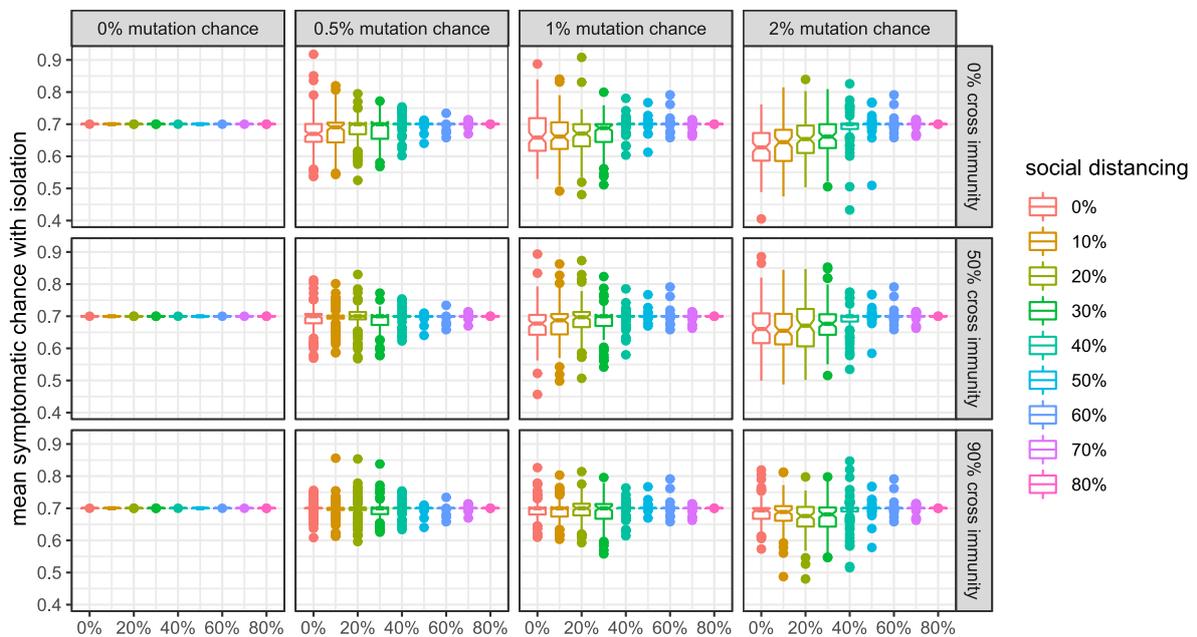

**Figure 19**: Mean probability of a symptomatic course of active (or, in case of extinction, last surviving) variants at simulation step 500 <u>with isolation of symptomatic individuals</u> and with 99% cross protection against a lethal infection (notched box plot).

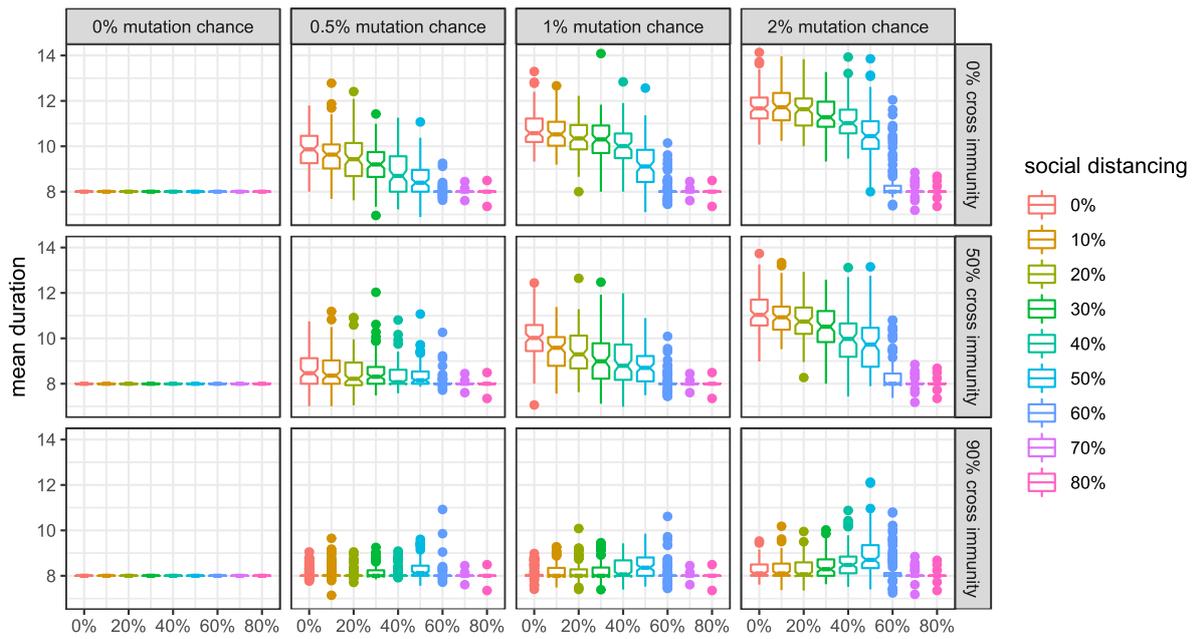

**Figure 20**: Mean duration of active (or, in case of extinction, last surviving) variants at simulation step 500 with 99% cross protection against a lethal infection (notched box plot).

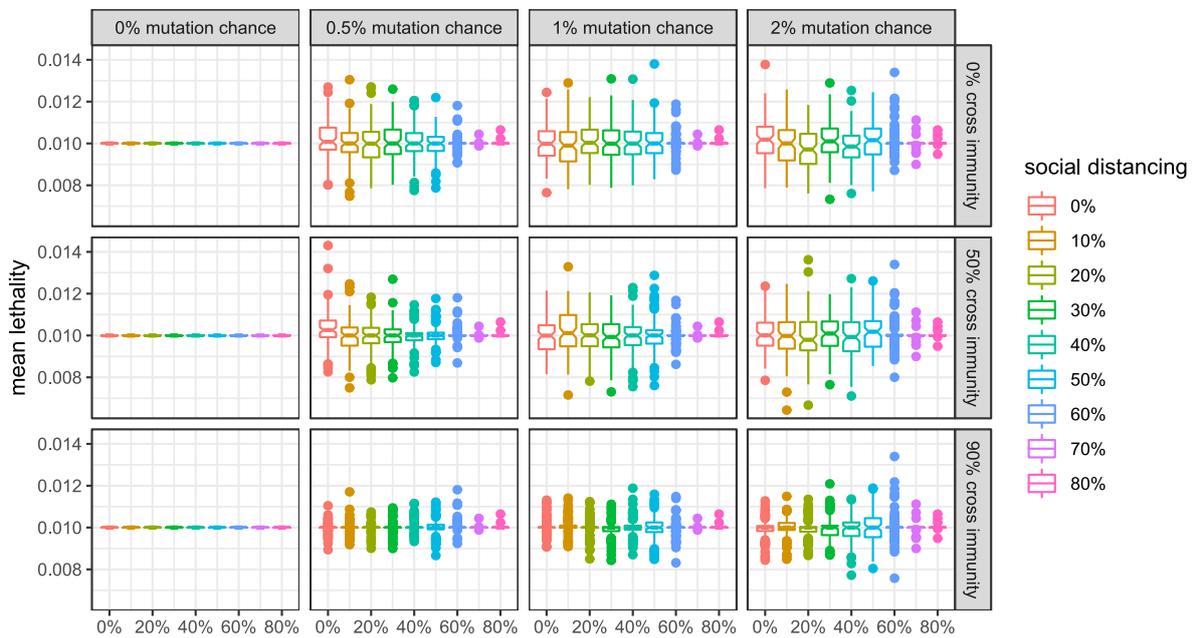

**Figure 21**: Mean lethality of active (or, in case of extinction, last surviving) variants at simulation step 500 with 99% cross protection against a lethal infection (notched box plot).